\documentclass[reprint,aps,prd,superscriptaddress,showpacs,preprintnumbers,amsmath,amssymb,floatfix]{revtex4-2} 

\usepackage[mathlines]{lineno}
\usepackage[mathlines]{lineno}
\usepackage{graphicx,subfigure}
\usepackage{epstopdf}
\usepackage{dcolumn}
\usepackage{bm}
\usepackage{color}
\usepackage{upgreek}
\usepackage{afterpage}
\usepackage{tikz, pgfplots}
\usepackage[version=4]{mhchem} 
\usepackage[colorlinks=true]{hyperref}
\usepackage{verbatim}

\begin{document}

\preprint{ \today}

\title{Ionization yield measurement in a germanium CDMSlite detector using photo-neutron sources}

\begin{abstract}

Two photo-neutron sources, $^{88}$Y$^{9}$Be and $^{124}$Sb$^{9}$Be, have been used to investigate the ionization yield of nuclear recoils in the CDMSlite germanium detectors by the SuperCDMS collaboration. This work evaluates the yield for nuclear recoil energies between 1\,keV and 7\,keV at a temperature of $\sim$\,50\,mK. We use a Geant4 simulation to model the neutron spectrum assuming a charge yield model that is a generalization of the standard Lindhard model and consists of two energy dependent parameters. We perform a likelihood analysis using the simulated neutron spectrum, modeled background, and experimental data to obtain the best fit values of the yield model. The ionization yield between recoil energies of 1\,keV and 7\,keV is shown to be significantly lower than predicted by the standard Lindhard model for germanium. There is a general lack of agreement among different experiments using a variety of techniques studying the low-energy range of the nuclear recoil yield, which is most critical for interpretation of direct dark matter searches. This suggests complexity in the physical process that many direct detection experiments use to model their primary signal detection mechanism and highlights the need for further studies to clarify underlying systematic effects that have not been well understood up to this point.

\end{abstract} 

\author{M.F.~Albakry} \affiliation{Department of Physics \& Astronomy, University of British Columbia, Vancouver, BC V6T 1Z1, Canada}\affiliation{TRIUMF, Vancouver, BC V6T 2A3, Canada}
\author{I.~Alkhatib} \affiliation{Department of Physics, University of Toronto, Toronto, ON M5S 1A7, Canada}
\author{D.W.P.~Amaral} \affiliation{Department of Physics, Durham University, Durham DH1 3LE, UK}
\author{T.~Aralis} \affiliation{Division of Physics, Mathematics, \& Astronomy, California Institute of Technology, Pasadena, CA 91125, USA}
\author{T.~Aramaki} \affiliation{Department of Physics, Northeastern University, 360 Huntington Avenue, Boston, MA 02115, USA}
\author{I.J.~Arnquist} \affiliation{Pacific Northwest National Laboratory, Richland, WA 99352, USA}
\author{I.~Ataee~Langroudy} \affiliation{Department of Physics and Astronomy, and the Mitchell Institute for Fundamental Physics and Astronomy, Texas A\&M University, College Station, TX 77843, USA}
\author{E.~Azadbakht} \affiliation{Department of Physics and Astronomy, and the Mitchell Institute for Fundamental Physics and Astronomy, Texas A\&M University, College Station, TX 77843, USA}
\author{S.~Banik} \affiliation{School of Physical Sciences, National Institute of Science Education and Research, Jatni - 752050, India}\affiliation{Homi Bhabha National Institute, Training School Complex, Anushaktinagar, Mumbai 400094, India}
\author{C.~Bathurst} \affiliation{Department of Physics, University of Florida, Gainesville, FL 32611, USA}
% ...

\smallskip
\date{\today}
\noaffiliation
\smallskip 
\maketitle

%
%
%
%
%
%
%
%

%
%
%
%
%
%
%

\section{\label{sec:intro}Introduction}
Many direct detection experiments search for dark matter particles that elastically scatter off the nucleus of a target material. The detectors themselves are sensitive to the recoiling nucleus and the dark matter interaction is inferred through detection of the nuclear recoil \cite{LEWIN199687}. The measured nuclear recoil energy ($E_r$) from such an interaction depends on the momentum transfer between the dark matter particle and the nucleus. A precise measurement of the recoil spectrum provides information about the mass of the putative dark matter particle. In semiconductor targets, such as germanium (Ge) or silicon (Si), the recoiling nucleus transfers its energy to ionization ($e^{-}/h^{+}$pairs) and phonons. The ratio of the ionization energy to the total recoil energy is commonly referred to as the ``ionization yield''. For detectors that rely on deriving the nuclear recoil energy from the ionization energy, a precise understanding of the ionization yield is needed~\cite{PhysRevD.95.082002}. A deeper knowledge of ionization yield is not only important for direct dark matter searches, but also for calibrating the nuclear recoil signal induced by coherent elastic neutrino nucleus scattering~\cite{Freedman-CENNS,Drukier-CENNS,COHERENT-CENNS}.

\begin{figure}[!ht]
    \centering
    \includegraphics[width=\linewidth]{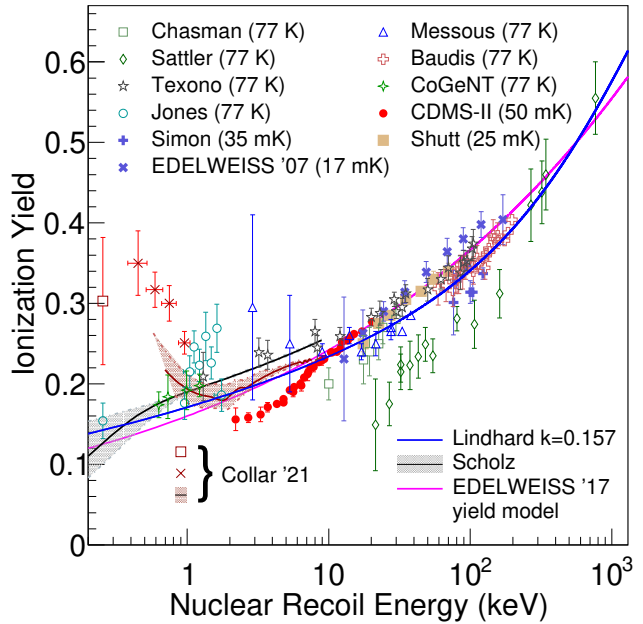}
    \caption{Literature data on ionization yield measurements as a function of recoil energy in germanium \cite{PhysRevC.4.125, PhysRevA.11.1347, Aalseth:2012if, Xichao, MESSOUS1995361, PhysRevLett.106.131302, MooreThesis, PhysRevLett.21.1430, PhysRevLett.69.3425, BENOIT2007558, PhysRev.143.588, BAUDIS1998348, SIMON2003643, Scholz:2016qos, Armengaud_2017, Collar21}. Measurements by Shutt, Simon, EDELWEISS, and CDMS-II are in the mK temperature range, while all others are at 77~K. The CDMS II yield measurements are from a single detector~\cite{PhysRevLett.106.131302} that is similar to the averaged yield over all detectors. However, variation across individual detectors was more than the statistical uncertainty of the measurement shown here~\cite{MooreThesis}. EDELWEISS has reported two measurements: (i) discrete yield measurements using a $^{252}$Cf neutron source~\cite{BENOIT2007558}, and (ii) continuous measurement using an $^{241}$Am$^{9}$Be neutron source fitted by a power law function considered as their yield model~\cite{Armengaud_2017}. The work by Collar\,{\it et al.} reports three different approaches to measuring ionization yield in germanium: (i) using photo-neutrons from an $^{88}$Y$^{9}$Be source indicated by the red band, (ii) using nuclear recoils from thermal neutron capture indicated by the red hollow square marker, and (iii) using an iron-filtered low energy neutron beam~\cite{Collar21}. The prediction by Lindhard et. al. \cite{osti_4153115, BARKER20121} is also shown for comparison.}
\label{fig:yield-all}
\end{figure}

Figure~\ref{fig:yield-all} shows the results from a number of experiments that have measured the ionization yield in Ge as a function of nuclear recoil energy. The figure also provides a comparison between the experimental results and a theoretical model of the ionization yield in Ge which is derived by Lindhard et al.~\cite{osti_4153115} and which is the reference model in the field of direct dark matter searches.  Experiments often use a value of the $k$ parameter (explained in Sec.~\ref{signal_model}), whose physical origin is related to the electronic stopping power~\cite{Horsfield}, to better match the Lindhard model to their experimental results. The Lindhard model, however, does not account for some factors that may be important at low energy, such as atomic binding~\cite{PhysRevA.11.1347}, electric field dependence~\cite{StanfordEField}, and temperature dependence~\cite{Wei_2017}. Thus the Lindhard model is not expected to be an accurate prediction of the ionization yield at the low recoil energies ($<$ 10 keV) expected from low mass dark matter particles ($< 5$\,GeV/c$^2$). Besides the shortcoming of the Lindhard model at low energies, it can also be noted from Fig.~\ref{fig:yield-all} that there exists some disagreement in the ionization yield measurements by various experiments in this recoil energy regime.

The SuperCDMS SNOLAB experiment will search for low mass dark matter using Ge and Si high-voltage (HV) detectors operated at temperatures below 40\,mK~\cite{PhysRevD.95.082002}. These detectors measure phonon signals that are proportional to the bias voltage and the amount of ionization in the detector due to a recoil event (see Sec.~\ref{signal_model} for details). The cryogenic HV detectors have a low recoil energy threshold, but cannot make a direct measurement of the ionization yield on an event-by-event basis as the ionization and phonon signals are measured together as a combined phonon energy response of the detector~\cite{PhysRevD.95.082002}. A key goal of this study is to provide an ionization yield measurement in Ge that is applicable to the SuperCDMS SNOLAB detectors at low recoil energies~\cite{PhysRevD.95.082002}.

In this work, we studied the nuclear recoil ionization yield in Ge using the SuperCDMS iZIP detectors~\cite{iZIP} in HV mode~\cite{PhysRevLett.112.041302} called CDMSlite~\cite{PhysRevLett.116.071301}, which were operated as part of the SuperCDMS Soudan experiment~\cite{Agnese:2018gze}. To generate nuclear recoils in the detectors, we exposed them to photo-neutron sources, which produce quasi-monoenergetic neutrons.  Such sources can be fabricated by pairing a radioactive source that has an intense, high energy gamma with Be wafers to induce photoproduction of neutrons from $^{9}$Be \cite{Wattenberg, RobinsonBe9}.  To extract the ionization yield, we performed a Geant4 simulation to model the recoil spectrum in the detectors.  We then performed a maximum likelihood fit using the input simulation spectrum and a two-parameter yield model to extract the ionization yield from the data.   We discuss the experimental setup and data sets in Section \ref{sec:experiment}, the data selection criteria and their selection efficiencies in Section \ref{sec:cuts}, the simulation of the nuclear recoil spectrum in Section \ref{sec:simulation}, the yield extraction using the likelihood function in Section \ref{sec:yield}, the results of this work are discussed in Section \ref{results}, and our concluding remarks are presented in Section \ref{sec:conclusions}.

\section{\label{sec:experiment}Experimental Setup}
\begin{figure}[h!]
    \centering
    \includegraphics[scale=0.501]{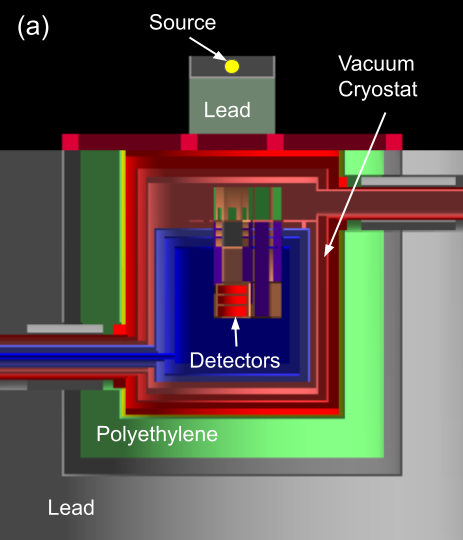}
    \includegraphics[scale=0.2]{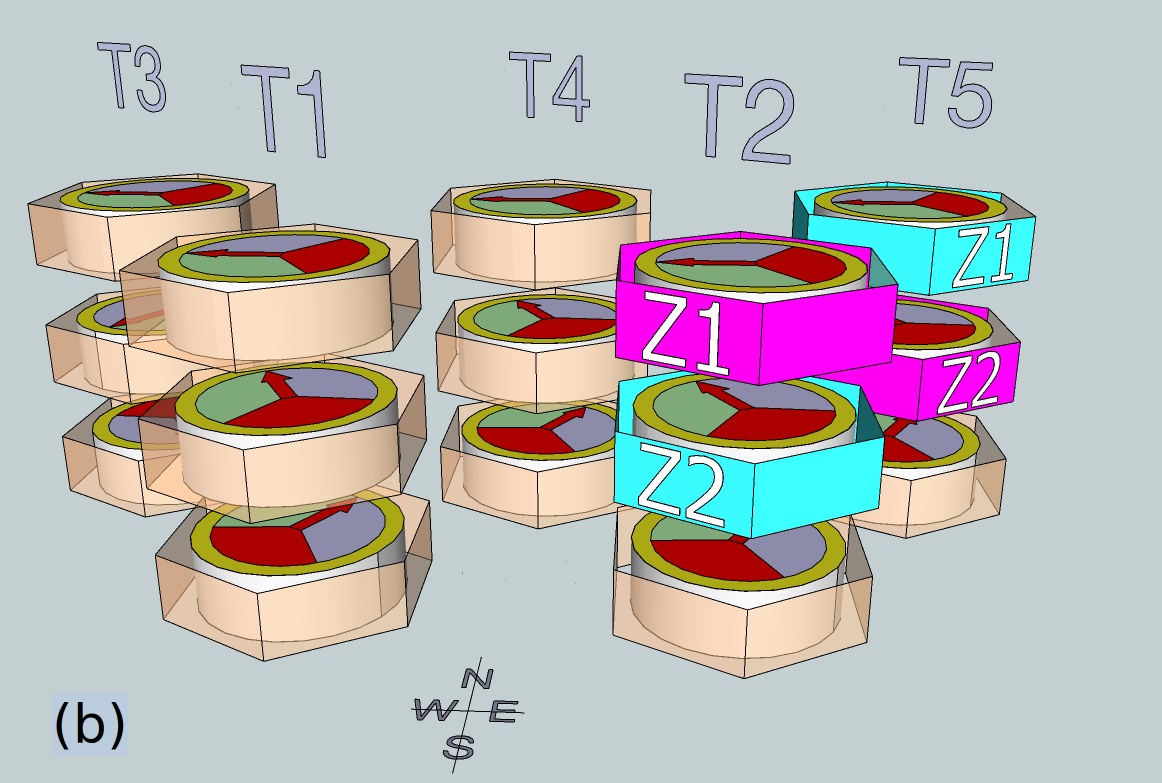}
    \caption{(a) The experimental setup with the shielding configuration. Lead was used to shield against gammas.  Polyethylene shielding was required for the dark matter configuration and only the top portion was removed from the experiment in order to perform the photo-neutron calibration. (b) Detectors were arranged into 5 towers with 3 detectors in each tower. Each face of the detectors had four phonon channels. The phonon channel layout is indicated by the different colours on the upper face of the detector. The photo-neutron source was placed above and between towers T2 and T5. The T2Z1 and T5Z2 detectors used in this analysis have been highlighted in pink. The detectors T2Z2 and T5Z1, highlighted in blue, were used for a systematic study of the event rate due to the position of the source over the two towers.}
    \label{fig:fig1}
\end{figure}

The SuperCDMS Soudan experiment, operated from 2012 to 2015, was located at the Soudan Underground Laboratory in northern Minnesota. The dark matter experiment consisted of 15 high-purity germanium detectors that were arranged in a compact array of five towers with three detectors per tower, as shown in Fig.~\ref{fig:fig1} (b). Each detector was roughly cylindrical in shape and approximately 600\,g in mass. The top and bottom faces of each detector were instrumented with tungsten transition-edge sensors (TES) coupled to aluminum energy-collecting ``fins" used to read-out athermal phonon signals and provide electrodes for measuring ionization signals. A dilution refrigerator cooled the payload to a baseline operating temperature that varied between 40 and 50 mK.  Further details about the experimental setup and the detector design can be found in Ref.~\cite{PhysRevLett.112.041302,CDMSlite2019}. 

To perform the photo-neutron calibration measurement, the experimental configuration was modified slightly after the main run that was used for dark matter searches.  Figure ~\ref{fig:fig1} (a) shows a schematic of the experiment as it was configured for the photo-neutron measurement.  The detectors were operated in two modes: (i) iZIP \cite{iZIP}, and (ii) CDMSlite \cite{CDMSlite}. In the CDMSlite mode, the detectors were biased at higher voltages (up to 70\,V) than in iZIP mode to take advantage of the Neganov-Trofimov-Luke (NTL) effect. The NTL effect is the production of secondary phonons in proportion to an ionization energy deposit and an applied bias voltage~\cite{Neganov,Luke}. The higher voltage applied to CDMSlite detectors thus served to amplify the phonon signal from the primary ionization, effectively lowering the energy threshold of the detector.  

Two of the detectors, labelled as T5Z2 and T2Z1 (see Fig.~\ref{fig:fig1} (b)), were operated in CDMSlite mode. They were biased at 70\,V and 25\,V, respectively. These detectors were chosen because they showed adequate signal to noise performance when biased at higher voltages. The photo-neutron source was placed above the respective towers (see Fig.~\ref{fig:fig1} (a)).

Two gamma sources, $^{88}$Y (half-life 106.6\,days) and $^{124}$Sb (half-life 60.2\,days), were deployed sequentially for the photo-neutron calibration. Each source was placed on top of a solid $^{9}$Be disk. $^{9}$Be is the only stable isotope of Be that occurs naturally in non-trace amounts. The activity of each of the gamma sources at the start of the calibration period was 1\,mCi. The most prominent gamma above the photo-disassociation threshold (1.66\,MeV) that is emitted by $^{88}$Y is at 1.8\,MeV.  In the case of $^{124}$Sb, the most prominent gamma above threshold is 1.69\,MeV. In the photoproduction process, the $^{9}$Be nucleus absorbs a gamma and emits a neutron to become a $^{8}$Be resonant state. The resulting neutrons emitted from $^{9}$Be, over multiple occurrences of this process, are nearly mono-energetic, with an average energy of 152\,keV \cite{Scholz:2016qos} for $^{88}$Y and 24\,keV \cite{PhysRevD.94.082007} for $^{124}$Sb. The diameter of the $^{9}$Be disc was 5\,cm with a thickness of 0.2\,cm. The purity level of the Be disc was 98.5\% $^{9}$Be. A source holder was made to constrain the position of the gamma source with respect to the $^{9}$Be disc.

There is a small spread in the energy of the emitted neutrons, which depends on the angle at which they were emitted with respect to the direction of the incident gamma~\cite{Wattenberg}. The spread in neutron energies for the $^{88}$Y$^{9}$Be source is $\sim8$ keV, and for the $^{124}$Sb$^{9}$Be source is $\sim1.3$ keV. Only neutrons predominantly moving in the forward direction (towards the detector) are likely to reach the detectors, due to the small solid angle subtended by the detector. The angle-energy correlation was not included in the Geant4 simulation; instead we used a mono-energetic energy consistent with the forward-direction. A cross-check was performed using the backward-direction energy, and the difference was assigned as a systematic uncertainty.

The ratio of all gammas to one neutron produced from the photo-neutron source is approximately $10^{5}$. Such a high rate of events had the potential to induce a prohibitive amount of dead-time for the data acquisition (DAQ). To suppress the high rate of these gammas, 13 to 15\,cm of lead shielding was placed between the photo-neutron source and the detectors at different stages of the run (see Fig.~\ref{fig:fig1} (a)). The thickness of the lead shielding between the source and the detectors was optimized to balance degradation of the neutron energy spectrum against the desire to adequately reduce the DAQ dead-time. Additionally, the thickness of the lead shielding between the photo-neutron source and the detectors was decreased as the decay rate of the source diminished.  A special ``window" trigger was implemented at the software level.  This served to further reduce the rate of stored events and manage the DAQ dead-time by vetoing high-energy electron recoils that produced energy deposits well above the nuclear recoil energy spectrum.

The photo-neutron data-taking took place for $\sim$144\,days between June 5th, 2015 and October 26th, 2015. Data were recorded with the $^{9}$Be disk (neutron-ON) and without the $^{9}$Be disk (neutron-OFF).  The latter served to measure the gamma background. A summary of the data-taking condition is shown in Table~\ref{table_rundetail}. The data was first taken with the T5Z2 detector, and subsequently with the T2Z1 detector. As the activity of the $^{124}$Sb source ($\tau_{1/2}$ = 60.2 days) had become much weaker by the time of T2Z1 operation in CDMSlite mode, no data were taken with this source in that operation period.

\begin{table}
 \caption{Summary of the photo-neutron data-taking runs.} 
\begin{ruledtabular}
\begin{tabular}{c c c c}
 Source  & Duration & Detector & Voltage\\
 $^{124}$Sb and $^{124}$Sb$^{9}$Be & 62 days & T5Z2 & 70 V\\
 $^{88}$Y and $^{88}$Y$^{9}$Be  & 42 days & T5Z2 & 70 V\\
 $^{88}$Y and $^{88}$Y$^{9}$Be  & 38 days & T2Z1 & 25 V\\
\end{tabular}
 \end{ruledtabular}
\label{table_rundetail}  
\end{table}

\section{\label{sec:cuts}Data reconstruction, selection and efficiencies}

The phonon pulse shape observed in the various channels features a steeply rising edge (few microseconds) and a slowly falling, exponential tail (with a time constant of a few milliseconds)~\cite{PhysRevD.97.022002}. The rising edge is governed by the timescale for the earliest phonons to reach the sensors and provides information about the position of the interaction. The phonon sensor coverage is relatively low ($\sim 5$\%), and thus the majority of phonons are reflected multiple times from uninstrumented surfaces before they are detected, contributing to the slowly falling tail. This slow component of the signal provides information about the total event energy and the tail shape is largely independent of position throughout the detector.

The data readout scheme and reconstruction are described in detail in Ref.~\cite{PhysRevD.97.022002}.  As described there, the energy estimation was done using a variation on an optimal filter (OF), referred to as the ``non-stationary" optimal filter (NSOF) which used a template to fit the recorded pulses. 
The template was made by averaging over a selection of recorded pulses that were deemed to be of high quality and representative of real, particle-induced events. The NSOF treats the differences between the template and the fitted pulse as non-stationary noise. Since variation in the rising edge of the pulse arises due to position dependence in the phonon response, the NSOF had the effect of deweighting the fit to position-dependent systematics and yielded a more precise energy estimation. In addition to fitting the recorded pulses to a standard pulse template, the pulses were also fit with templates that matched ``glitches'' and typical low frequency noise (LFN). The glitches are electronic in origin and are characterised by pulses with an unusually fast rise and fall time. The LFN ($<$ 1\,kHz) originated from vibrations created by a cryocooler, which provided supplemental cooling power to the experiment. A correlation between the cryocooler activity and the LFN in CDMSlite detectors was observed and reported in detail in Ref.~\cite{PhysRevD.97.022002}. The $\chi^2$ goodness of fit from the NSOF was stored for each fit and used for the selection of events as described in more detail below.

Prior to determining the ionization yield for nuclear recoils, it was necessary to calibrate the response of the detectors to electron recoils. To do this, the germanium detectors were exposed to neutrons from $^{252}$Cf. This led to a neutron capture process, creating unstable $^{71}$Ge. $^{71}$Ge decays via electron capture to $^{71}$Ga, releasing predominantly x-rays with a total energy corresponding to the binding energy of K-, L- or M-shell electrons. The energy scale of each detector was calibrated using the 10.3\,keV and 1.3\,keV peaks from the K- and L-shell electron capture lines in $^{71}$Ge \cite{PhysRevD.97.022002, Hampel, Bearden}. The calibration values from a prior dark matter search were adopted for this work, following the above described technique, and are described in further detail in Ref.~\cite{PhysRevD.97.022002}. For this dataset, the stability of the adopted calibration values was then checked using the electron capture lines as observed in dedicated low background datasets (i.e. no source placed during the run) taken at periodic intervals before, after, and between the photo-neutron runs. These checks showed that the energy calibration was stable to within 2\% throughout  the several-months period when photo-neutron data was collected.

To optimize data quality, several event selection criteria (cuts) were applied in the analysis. A ``livetime" cut was used to remove specific time periods of data from the analysis. ``Quality cuts" were applied on an event-by-event basis using the recorded pulse shapes. A ``threshold" cut was used to set the low-energy analysis threshold.

\textbf{\emph{Livetime cuts:}} Cuts were applied to ensure data were taken with the correct running conditions.  These included checks on the base temperature, detector bias voltage, and time between LED flashes (near-IR LEDs were used to release trapped charges from the detector) \cite{Chagani:2012zz}. Additionally $\sim1$\% of livetime was removed during periods of unusually high event rates, based on the average trigger rate calculated every 30 s of runtime. A few runs (totaling less than 12 hours) that showed an exceptionally low noise environment were also removed from the dataset. The ``pre-trigger" baseline for each phonon pulse was recorded for approximately 1 ms before each event. Excessive fluctuation or large RMS in the pre-trigger baseline is indicative of noise or event pileup. A cut was applied to remove events for which the standard deviation in baseline of the phonon pre-trigger region deviated by more than $4\sigma$ from its mean value, as calculated over the course of the run. The livetime cuts removed $\sim$10\% of the events from the 70\,V datasets, and $\sim$5\% of the events from the 25\,V dataset.

\textbf{\emph{Quality cuts:}} 
As described above, the energy of each event was reconstructed from the phonon pulses using a variant on an optimal filter routine called NSOF.  Using the NSOF template fit to phonon pulses, a $\chi^{2}$ goodness of fit was calculated for each event. The $\chi^{2}$ as a function of fit energy was constant over the neutron scattering energy range, indicating that the fit was well-behaved over the energy range of interest. Events with very large $\chi^{2}$ or with a rise time outside a 230\,$\mu$s window around the trigger time were removed from the analysis. Such events typically consisted of pileup events or other poorly reconstructed events. 

In the CDMSlite detectors, the signals from the ionization channels cannot be used in the reconstruction of the photo-neutron recoil spectrum, as they do not recover the entire charge energy from an ER or NR event. However, positive correlations were found between the charge and phonon goodness of fit for extremely high charge OF $\chi^{2}$ values ($\chi^{2}>$6250 for data from T2Z1, and $\chi^{2}>$50000 for data from T5Z2). Thus, events with a high charge $\chi^{2}$, which typically displayed a sharp glitch in the waveform, were rejected from the analysis to remove bad events that survived phonon-based selection criteria. 

In addition to the standard pulse template, each phonon pulse was also fit with both an LFN and a glitch template. Based on the goodness of the respective fits, a variable discriminating between signal-like and noise-like events called $\Delta OF\chi^2_{\rm noise} = OF\chi^2_{\rm pulse}-OF\chi^2_{\rm noise}$ was constructed, where the subscript ``noise" stands for either LFN or glitch.  A cut at an appropriate level of signal acceptance (3$\sigma$) was applied on the $\Delta OF\chi^2_{\rm noise}$ distribution as a function of total phonon energy to remove both classes of noise events. To best optimize the cuts, the photo-neutron dataset was divided into eleven time blocks such that the behavior of the LFN associated with the cryocooler within each time block was similar. The cuts to reject LFN and glitches were defined for each of the time blocks separately.

In the data acquisition system, charge triggers were recorded when the charge pulse exceeded a discriminator threshold.  Only phonon triggers were used to make a decision on recording the data, however the charge triggers were recorded and used for data quality studies. A good pulse would ideally issue a trigger in both the charge and phonon channels, but glitches on a particular channel could create a difference in the number of triggers issued between the channels. Thus, glitch events were further rejected by requiring that selected events had coincident triggers from both the phonon and charge channels. 

\begin{figure}[!htb]
    \includegraphics[width=\columnwidth]{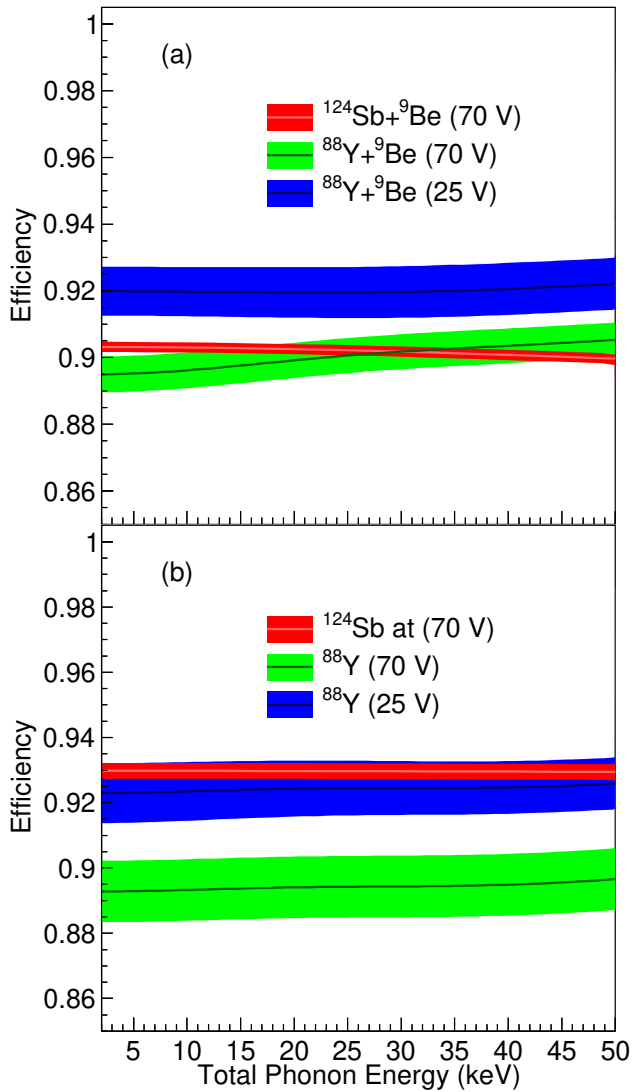}
    \caption{\label{fig:cut_efficiency} Combined and smoothed cut efficiency as a function of energy above the analysis threshold for each data set with a $1 \sigma$ uncertainty. (a) With the Be wafer i.e. neutron-ON, and (b) without the Be wafer i.e. neutron-OFF. }
\end{figure}

\textbf{\emph{Threshold cut:}} 
For a given detector, the phonon trigger efficiency calculation relied on events triggered by the two other detectors in the same tower. Only data from the triggering detector were read out in the photo-neutron data-taking mode, hence a separate set of data had to be utilized to measure the trigger efficiency.  These data required that an entire tower of detectors be read out whenever a single detector passed the phonon trigger condition. This general technique has been used by SuperCDMS in the past and is described in references~\cite{PhysRevD.97.022002}~\cite{CDMS:2010ntd}.  The trigger efficiency for this analysis is based on a combination of data taken with a $^{252}$Cf source and data taken in the low background mode.  These datasets were selected because they had trigger thresholds matching those in the photo-neutron runs.  They also provided data with a continuous neutron energy distribution up to a few MeV and real physical events that induced multiple triggering events in the full tower of detectors. To minimize the potential impact of the systematic uncertainty in the trigger efficiency calculation on our results, the analysis energy thresholds were set to just above where the trigger efficiency was uniformly 100\%. Lastly, in order to interpret  the threshold values in terms of recoil energy we applied a correction for the NTL phonons. For the case of electron recoils (yield$\equiv$1), the threshold for this analysis was  74\,eV for T5Z2 and 236\,eV for T2Z1.

We calculated the efficiencies of the quality-cuts described previously by relying on a data driven approach following the procedures described in Ref.~\cite{CDMSlite2019}. To compute the efficiencies, we generated artificial events by adding a scaled pulse template to random noise traces gathered during the data acquisition. To take position dependencies into account the pulse templates were generated as a linear combination of two sub-templates modeling the fast and slow components of the absorption signal described above and in more detail in Ref.~\cite{PhysRevD.97.022002}. The relative weights of the sub-templates were drawn from the fast-to-slow amplitude ratio distribution observed in the photo-neutron data. These artificial data, generated for each data set in a time block, reproduced similar distributions as observed for the measured total phonon energy and the phonon OF $\chi^{2}$ in the real data. The efficiency for each time block was calculated as the ratio of the events that pass the cuts over the total number of generated events, as a function of energy. For the final calculation, the efficiency functions of the constituent time periods were combined together, weighting each period by both its livetime and the decreasing activity of the source (see Sec.~\ref{sec:yield}).  The efficiency functions were then smoothed with a Gaussian filter to reduce  the effects of statistical fluctuations. 
Figure~\ref{fig:cut_efficiency} shows the resulting combined and smoothed cut efficiency functions for each data set. As seen in the figure, the quality cuts were generally above 90\% efficient and showed relatively little energy dependence.   

Between data sets, where the source configuration was changed, we observed discrete jumps down in rate by as much as 5\%, which is not compatible with statistical fluctuations in the rate or known changes in operating conditions. The same  jumps were present in the data when no selection cuts were applied. We also measured a smooth downward trend in rate that was consistent with the exponential decay of the sources. A systematic study on the effect of a slight deviation in the source position on the overall event rate revealed that this effect was negligible and not consistent with the rate jumps.

We used a $\chi^2$ test to determine whether the jumps in rate affected the shape of the neutron-ON energy distribution. To test the consistency of the energy spectra over time, a $\chi^2$ was calculated between pairs of data sets drawn from the eleven time periods and the same photo-neutron source. The $\chi^2$ tests returned a p-value which informed whether the data were drawn from the same parent distribution.  Out of the 62 tests we performed, only one returned a p-value lower than 0.01.  Based on this result, we concluded that the shape of the energy spectrum is consistent over time and the fit should not be affected by the rate fluctuations seen between data sets.

\section{\label{sec:simulation}Simulations}

\begin{figure}[!ht]
    \includegraphics[width=\columnwidth]{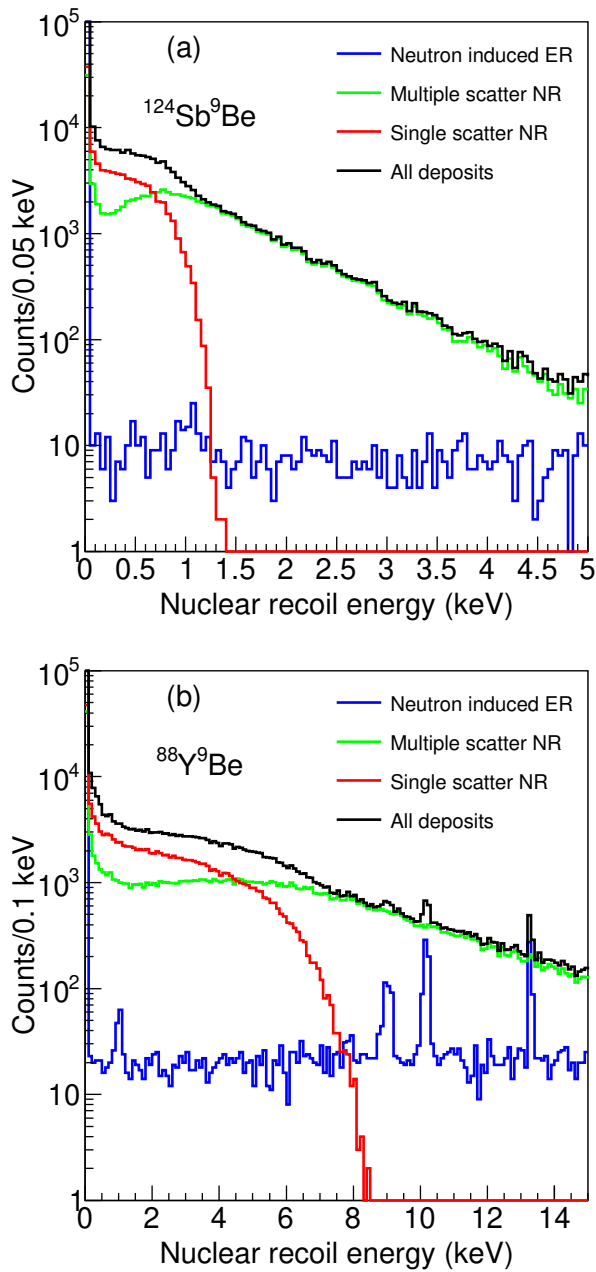}
    \caption{\label{fig:singles_multiples} The simulated recoil energy distributions from (a) $^{124}$Sb$^{9}$Be, and (b) $^{88}$Y$^{9}$Be in T5Z2. The energy depositions from electron recoils (ER) that arise from secondary gammas is shown in blue, the energy depositions from multiple-scatter nuclear recoils (NR) by neutrons in the detector is shown in green. The energy distribution of single-scatter NR by neutrons is shown in red.}
\end{figure}

A primary input to the likelihood calculation is the neutron energy probability distribution function (PDF). Geant4 10.6 \cite{GEANT4} was used to simulate the neutron energy spectrum from the different source configurations. The Geant4 NeutronHP physics model~\cite{NeutronHP} along with the G4NDL4.6 cross-section packages were implemented for the simulation of 1.2$\times 10^9$  neutrons propagating through the experimental geometry. G4NDL4.6 is the JEFF3.3 library. It produces results closer to MCNP, has more isotopes than ENDF/B, and has an overall lower error in the energies of secondary neutrons~\cite{G4NDL4.6-validation,JEFF33}. The Compton spectrum from the source gammas was modeled with an analytical spectrum and cross-checked with the neutron-OFF data, as described in detail in Sec.~\ref{bkg_model}. Therefore, the only simulated data used in this analysis was the neutron sample, which included a very small contribution of secondary gammas originating from neutron interactions in the surrounding materials.  

Figure \ref{fig:singles_multiples} shows the recoil energy deposition in T5Z2 of neutrons from the $^{124}$Sb$^{9}$Be and $^{88}$Y$^{9}$Be sources in panels (a) and (b) respectively. Single and multiple scatter nuclear recoil (NR) energy deposits from neutrons interacting directly in the detector and electron recoils (ER) from neutron-induced gammas are shown separately. The sharp drop-off in the single-scatter neutron spectrum (``endpoint") corresponds to the maximum energy a neutron of the given input energy can transfer to a Ge nucleus in a single interaction ($\sim$1.3 keV for the 24 keV neutrons from $^{124}$Sb$^{9}$Be and $\sim$8.2 keV for the 152 keV neutrons from $^{88}$Y$^{9}$Be; see Fig. \ref{fig:singles_multiples}).

There are a large number of multiple neutron scatters, which obscure the endpoint of the single scatter neutrons in the total energy distribution. This makes it difficult to determine the exact position of the maximum recoil for single-scatters in the actual data distribution, and thus to directly determine the ionization yield with an integral-based method as described in Ref.~\cite{PhysRevD.94.082007}. Additionally, the neutrons pass through layers of lead, polyethylene, and copper before reaching the germanium detectors. This can produce neutron induced gammas from these materials giving rise to an ER background. As seen in Fig.~\ref{fig:singles_multiples}, the rate of the ER background is low compared to the neutron scattering rate and is almost flat over the entire energy range. Furthermore it is subdominant to the Compton scattering spectrum that results from the gammas directly produced by the source (as described in the next section). The various peaks seen in Fig.~\ref{fig:singles_multiples}\,(b) arise from the electron capture processes in germanium. 

\begin{figure}[!h]
    \includegraphics[width=\columnwidth]{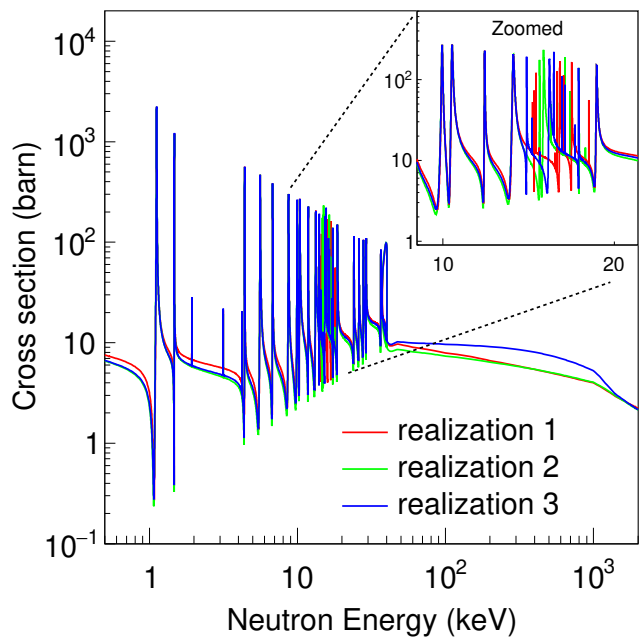}
    \caption{\label{fig:crosssections} Three examples of realizations of neutron-nucleus scattering cross-section over a large neutron energy range from the TENDL-2017 nuclear data library~\cite{TALYS}. Inset is zoomed into the neutron energy range of interest where the differences between the cross sections are considerable between the three realizations. See text body for description of how TENDL cross-section realizations are generated.}
\end{figure}

We studied two systematic uncertainties in the neutron PDF arising from the simulation. The first originates from the choice of physics model used in Geant4. To study this uncertainty, the simulation data were regenerated with the same sample size as used for the primary analysis with the LEND~\cite{LEND} package in place of the NeutronHP~\cite{NeutronHP} and G4NDL4.6 packages~\cite{G4NDL46}. The likelihood analysis was re-done using this alternative simulation dataset (see Sec.~\ref{Uncertainties}). The second uncertainty is introduced by the neutron-nucleus scattering cross-section information available for germanium through the Geant4 package. Figure~\ref{fig:crosssections} shows the cross-section in germanium as a function of neutron energy for three different example realizations of germanium cross-section files from the TENDL-2017 nuclear data library \cite{TALYS}. The TENDL data library is created via the Total Monte Carlo method \cite{MC-TENDL}, wherein input parameters to the TALYS nuclear reaction simulation are randomly varied around known central values, which are in turn determined from a combination of experimental data and computational models. The resulting cross-sections show some differences, particularly in the neutron energy region below 50\,keV.  Using NeutronHP, we re-ran the simulation with 100 different realizations of the neutron cross-sections. Each simulation was for 6$\times 10^8$ neutrons. These simulation datasets were used for determining the systematic uncertainty introduced by this effect as described further in Sec.~\ref{Uncertainties}. The two systematics are comparable in size, with the one from using the LEND library instead of the NeutronHP and G4NDL4.6 being slightly larger.

\section{\label{sec:yield}Yield extraction}
In order to extract the nuclear recoil ionization yield, we parameterized the conversion from total phonon energy to nuclear recoil energy following a Lindhard-like model where the parameters were determined by fitting the model to the neutron-ON data. The fit minimized the summed negative log likelihood of all events given a PDF, which is the sum of a PDF of the gamma background model and a PDF for the neutron component. The neutron component is constructed from the values of the yield modeling parameters, the simulated neutron spectrum, and a model of the detector response. The signal modeling is described in Sec.~\ref{signal_model}, and the background model in Sec.~\ref{bkg_model}. Section~\ref{likelihood} discusses the details of the likelihood fit, the propagation of statistical and systematic uncertainties, and the final results.

\subsection{\label{signal_model}Signal modeling}

The recoil energy ($E_r$) distributions from the two photo-neutron sources were derived from the simulated neutron data described in Sec.~\ref{sec:simulation}. In order to compare the simulation to real data, the simulated energy deposition was converted into the total phonon energy ($E_p$), the observable measured by the detectors. 

The neutron signal is modeled for the three source configurations: (i) $^{124}$Sb$^{9}$Be with the detector biased at 70\,V, (ii) $^{88}$Y$^{9}$Be with the detector biased at 70\,V, and (iii) $^{88}$Y$^{9}$Be with the detector biased at 25\,V. When a bias voltage is applied to the detectors, the phonon energy generated through the NTL effect must be taken into account. Work done by the electric field ($E_{\text{NTL}}$) in drifting the $e^{-}/h^{+}$pairs to the charge electrodes is converted into NTL phonons which are detected by the TES sensors along with the phonons generated by the primary recoil interaction. The total phonon energy is the sum of the recoil and NTL energies. The NTL energy is calculated as
\begin{equation}
 E_{\text{NTL}}=N_{e/h} e \Delta V=Y(E_r)\frac{E_r}{\epsilon_{\gamma}}e \Delta V,
\end{equation}
where $N_{e/h}$ is the number of $e^{-}/h^{+}$pairs created by the recoil, $\Delta V$ is the bias voltage, and $e$ is the electronic charge. $Y(E_r)$ is the ionization yield, defined as the ratio between the ionization signal and the primary recoil energy, and $\epsilon_{\gamma}=3$\,eV~\cite{ANTMAN1966272, PEHL196845, Emery} is the average energy required per generated $e^{-}/h^{+}$pair in an ER interaction in germanium.

The total phonon energy can now be written as,
\begin{equation}
E_p = E_r(1+ Y(E_r) e \Delta V/\epsilon_{\gamma}),
\end{equation}
where the dependence of $E_p$ on the parameter of 
interest, $Y(E_r)$, appears clearly. 

The Lindhard theory~\cite{PhysRev.124.128, osti_4153115, lindhard1968approximation, osti_4701226}, supported by earlier measurements~\cite{ BARKER20121}, provides a semi-empirical prediction for $Y(E_r)$ of NR interactions for a material of mass number $A$ and atomic number $Z$:
\begin{equation}
 Y(E_r) = k\frac{g(\epsilon)}{1+kg(\epsilon)},
 \label{eq:yield}
\end{equation}
where: 
\begin{equation}
 g(\epsilon)  = 3\epsilon^{0.15} + 0.7\epsilon^{0.6}+\epsilon; ~~
 \epsilon  = 11.5\, E_r\,Z^{-7/3}.
 \end{equation}
The parameter $k$ describes the electronic energy loss and nominally ranges from 0.156 to 0.160 for stable isotopes 
of Ge. 

We introduce a modification to the Lindhard model in which we allow the $k$ value to vary linearly with the recoil energy:
\begin{equation}
 k(E_r) = k_\text{{low}}+\frac{ k_{\text{high}}-k_{\text{low}}}{E_{\text{high}}-E_{\text{low}}}(E_r - E_{\text{low}}).
\label{eq:kparameters}
\end{equation}

$E_{\text{low}}$ is equal to 0.39\,keV  and $E_{\text{high}}$ is equal to 7.0\,keV. These are the minimum and maximum nuclear recoil energy that the fit was sensitive to. In the following, we refer to $k_{\text{low}}$ and $k_{\text{high}}$ as the two components of a vector $\vec{k}$. 

In order to compare the simulated neutron signal to the collected data, we included effects due to the detector resolution to the simulated neutron spectrum. The resolution model ($\sigma_T$) used was:
\begin{equation}
\sigma_T^2 (E) = \sigma_B^2 + \sigma_F^2 (E) + \sigma_{D}^2 (E).
\label{eq:resolution}
\end{equation}
where $\sigma_B$ is the baseline noise due to the electronics, $\sigma_F$ is the variance in the number of $e^{-}/h^{+}$pairs produced in a recoil event, and $\sigma_D$ is an empirical term that models detector effects and is given by $\sigma_{D}^2 = AE^2$. 

The values of $\sigma_B$ are obtained from the energy distribution of events randomly triggered in the detector and are listed in Tab.~\ref{table_resolution}.  The second term can be written as $\sigma_F^2 =F N_{e/h}\epsilon_\gamma^2 = F \epsilon_\gamma E$, where $F$ is the Fano factor. The resolution model can thus be rewritten as:

\begin{equation}
\sigma_T (E) =\sqrt{ \sigma_B^2 + F\epsilon_\gamma E + AE^2},
\label{eq:resolution2}
\end{equation}

To determine the values of $F$ and $A$ in Eq.~\ref{eq:resolution2}, the resolutions of each of the K-, L- and M-shell peaks from electron-captures (EC) decays of $^{71}$Ge were determined by first fitting them with a Gaussian and obtaining their 1$\sigma$ values. Next, the model described in Eq.~\ref{eq:resolution2} was fit to the resolutions of these EC peaks. The values of $F$ and $A$ were determined from this fit. Table ~\ref{table_resolution} shows the values of the different parameters determined for the resolution model at the operating voltages of 70 V and 25 V. Since the Fano factor for nuclear recoils vs electron recoils can be significantly different, a systematic uncertainty on F is discussed in~\ref{Uncertainties}. 

\begin{table}
 \caption{Values of the parameters derived from the energy resolution model. The energy scale is electron-equivalent (eV$_{ee}$), where all events are assumed to be electron-recoils.} 
\begin{ruledtabular}
\begin{tabular}{c c c c} 
  Bias voltage & $\sigma_B$ (eV$_{ee}$) & $F$ & $A$\\
 70\,V & 8.31$\pm$0.08 & 0.19$\pm$0.07 & 0.0095$\pm$0.0015\\
 25\,V & 18.71$\pm$0.16 & 0.27$\pm$0.02 & 0.0107$\pm$0.0005\\
\end{tabular}
 \end{ruledtabular}
\label{table_resolution}  
\end{table}

Spatial variations in the uniformity of the electric field caused a variation in the amount of NTL gain achieved. Near the edge of the detector, the electric field is inhomogeneous, leading to a reduction in the observed NTL phonon contribution. To account for this a 3-D electric field map was calculated and taken into account when determining the total energy of the simulated events. Details of this technique are described in Ref.~\cite{PhysRevD.97.022002}.

Finally, to determine the neutron PDFs used in fitting the data, the resolution-smeared energies were converted to the total phonon energy using the parametrized model of Eq.~\ref{eq:kparameters}. In order to mitigate the statistical fluctuation effects, we smoothed the energy distributions using a Savitzky-Golay filter~\cite{SG_filter} before interpolating them using a cubic spline to determine the PDF.

\begin{figure*}[!htb]
  \includegraphics[width=\textwidth]{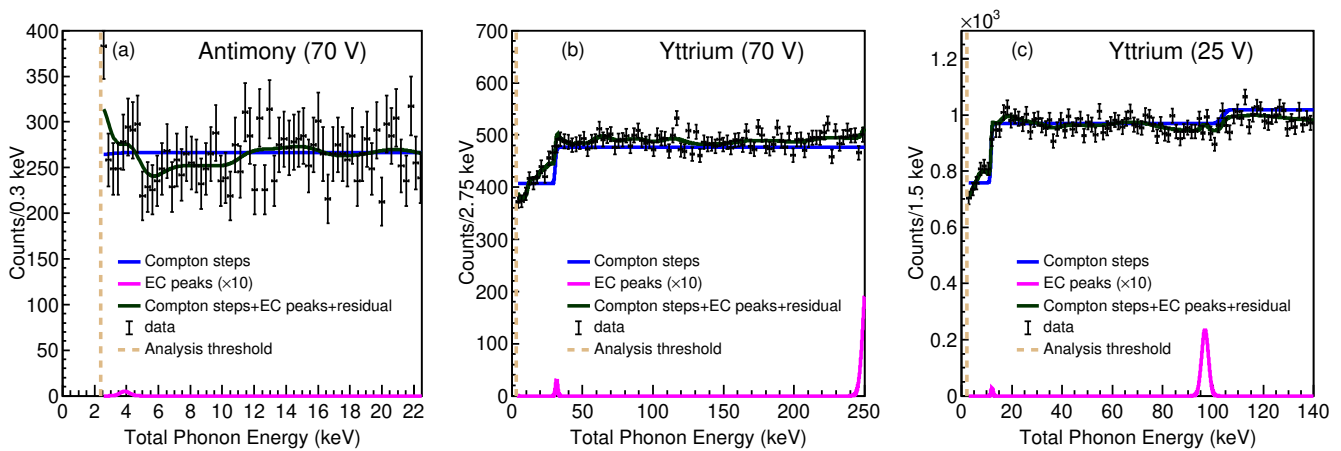}
  \caption{\label{fig:bkg_fit}Neutron-OFF energy spectrum for the three data sets: (a) $^{124}$Sb$^{9}$Be with T5Z2 biased at 70\,V, (b) $^{88}$Y$^{9}$Be with T5Z2 biased at 70\,V and (c) $^{88}$Y$^{9}$Be with T2Z1 biased at 25\,V. The contribution from each component in the background model is shown. The contribution of the electron capture peaks have been scaled up by a factor of ten for better visualization.}
\end{figure*}

\subsection{Modeling electron recoil backgrounds}
\label{bkg_model}

The dominant background that affected this measurement was photons from the source Compton-scattering off the electrons in the Ge crystal. Another source of background was the K-, L- and M-shell electron capture x-rays.  

The Compton continuum is characterized by steps in the scattering rate (i.e., x-ray absorption edges)~\cite{Ramanathan:2017dfn}, corresponding to each of the Ge electron shells. The Compton model in the region of interest is:
 \begin{equation}
 \begin{aligned}
  f_C(E) = a_0 [1 &+ S_M \Theta(E, \mu_M)]\\ +  S_La_0[1&+S_M \Theta(E,\mu_M)]\Theta(E,\mu_L)\\ + S_K a_0  [1 &+ S_M \Theta (E, \mu_M )]\Theta (E, \mu_L ) \Theta (E, \mu_K);
  \end{aligned}
 \end{equation}
 here $S_i$ is the fractional step amplitude where $i$ can be the K-, L-, or M-shell, and $a_0$ is the flat background level. The step sizes and background level were free parameters that were determined by fitting the neutron-OFF data. The parameter $\mu$ is the central energy 
 of the step, which was determined via a {\sc Geant4} simulation. $\Theta$ is the error function: 
 \begin{equation}
   \Theta(E, \mu)= \frac{1}{2}\left[1+ \text{erf}\left(\frac{E- \mu}{\sqrt{2} \sigma(\mu)}\right)\right],
 \end{equation}
 with the detector resolution $\sigma(\mu)$.

 The electron capture peaks were modeled using Gaussian functions. Following Ref.~\cite{SCHONFELD19981353}, we assumed the fractional electron capture probabilities for each shell to be $f_K = 87.6\%$, $f_L = 10.5\%$, and $f_M = 1.8\%$ for the K-, L-, and M-shells, respectively. The only parameter determined with a fit to the neutron-OFF data for this background was the amplitude of the K-shell peak. We then constrained the other amplitudes using the fractional ratio. While data collected at 70\,V are sensitive to the L-, K-, and M-shells, for the 25\,V data the M-shell step occurs below the analysis threshold, and was thus excluded from the model of the two backgrounds.

The combined background model from the EC peaks and Compton steps was fit to the neutron-OFF energy distribution. In order to include any outstanding effect which is not modeled, we added the fit residual, smoothed by applying a  Gaussian filter, to the model. Fig.~\ref{fig:bkg_fit} shows the neutron-OFF energy spectrum with the background model overlaid. The Compton steps are clearly visible and described by the model for the Yttrium at 70 and 25\,V. A systematic uncertainty for the background model is presented in Sec.~\ref{Uncertainties}.

Finally, the energy-dependent cut efficiencies for the quality cuts described in Sec.~\ref{sec:cuts} were applied to the signal and background PDF in order to represent the expected shape of the experimental data with all cuts applied. For the background PDF, since it was obtained by exploiting the neutron-OFF data, we first applied the inverse of the neutron-OFF cut efficiency function to remove any effects on its shape due to the neutron-OFF cut efficiency.

\subsection{Likelihood analysis}
\label{likelihood}

\begin{figure}[!htp]
    \includegraphics[width=0.8\columnwidth]{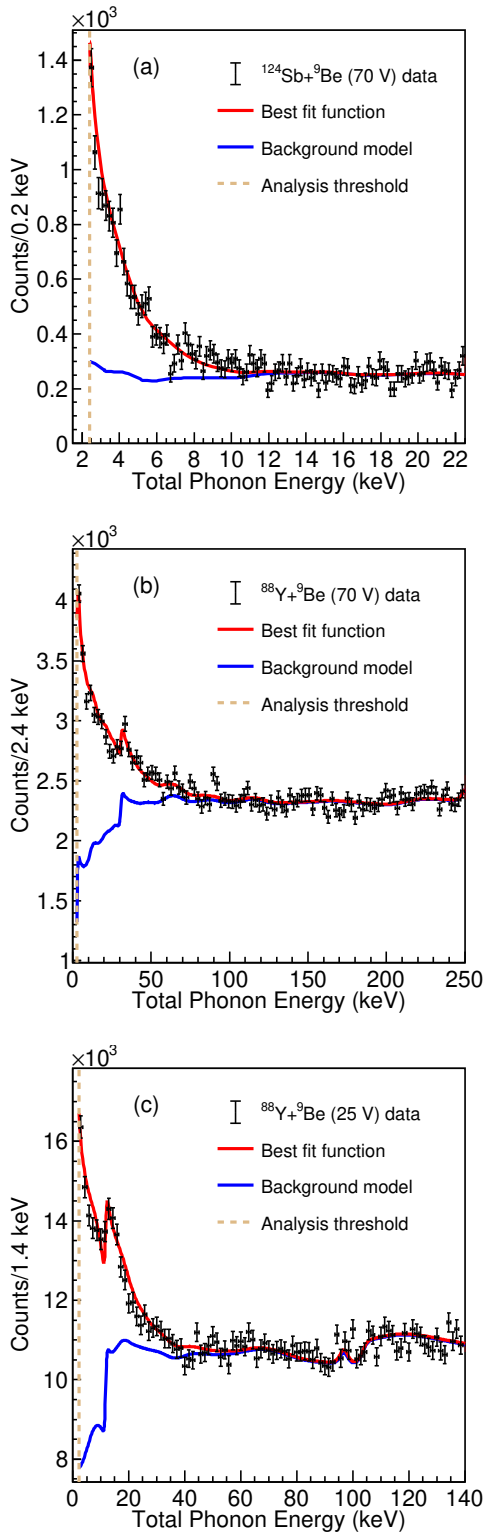}
    \caption{\label{fig:data_fit} Energy spectrum for the three data sets, (a) $^{124}$Sb$^{9}$Be with T5Z2 biased at 70\,V, (b) $^{88}$Y$^{9}$Be with T5Z2 biased at 70\,V and (c) $^{88}$Y$^{9}$Be with T2Z1 biased at 25\,V. The best fit result for each dataset is overlaid in red. The blue curve shows the fit gamma background contribution.}
\end{figure}

The background and signal PDFs allowed us to define the negative log likelihood function as:
\begin{equation}
 -\text{ln}\, \mathcal{L} = - \sum_{D=1}^3 \sum_{i=1}^{N_D} \text{ln}( f_D \nu_D(E_i, \vec{k}) + (1-f_D)b_D (E_i) ),
\end{equation}
where $N_D$ is the number of events in the data set $D$, $f_D$ is the fractional contribution of the neutron signal, $\nu_D (E, \vec{k})$ are the parameter-dependent signal PDFs, and $b_D (E)$ are the background PDFs. The free parameters of the negative log likelihood function, which were minimized using the MINUIT~\cite{MINUIT} package, were the three neutron contribution fractions ${f}_D$ and the Lindhard parameters $\vec{k}$. The relative goodness of fit, which takes into account only the statistical uncertainty, was evaluated by binning the experimental data and calculating the fit $\chi^2$ per degree of freedom for each data set. Figure~\ref{fig:data_fit} shows the energy spectra with the best fit result, for the 2-parameter model, overlaid on the three data sets. The $\chi^2$/\textit{(degrees of freedom)} was found to be $137/95$, $141/95$ and $129/95$ for $^{124}$Sb at 70\,V, $^{88}$Y at 70\,V and $^{88}$Y at 25\,V respectively. The best fit values of $\vec{k}$ are shown in Tab.~\ref{table_fitresults}.

In order to study whether the 1-parameter (standard Lindhard) or 2-parameter (see Eq.\,\ref{eq:kparameters}) $k$ model better described the data, we performed a likelihood ratio test. We generated 3000 Monte Carlo (MC) data sets sampling a number of events equal to the experimental data, from the background model plus the simulated neutron spectrum, which was converted to total phonon energy using the 1-parameter Lindhard, see Eq.\,\ref{eq:yield}. The ratio between the number of background and neutron events and the value of the $k$ parameter in the MC data generation were set to the best-fit values. We fit the MC data both using the 1-parameter and the 2-parameter modified Lindhard model. The likelihoods obtained in the two fits were used to construct the likelihood ratio, which was used as a test statistic to determine which model was a better fit to the data. Since the two models were nested, we used the likelihood ratio test to compare the goodness of fits in the two cases.  Given that the 1-parameter model was used to generate the MC data, the distribution of the likelihood ratio obtained with the MC data samples provides a quantitative measurement of how much better the 2-parameter model fit performed compared to the 1-parameter model, based on only statistical fluctuations. 

Finally, we calculated the likelihood ratio using the measured data and we compared it to the distribution obtained using the MC data samples.  We found no single likelihood ratio value calculated using the simulated data that had a value greater than the collected data.  We thus inferred that the 2-parameter hypothesis, which allowed for a linear energy dependence of the $k$ parameter of the Lindhard model, was preferred with a significance greater than 3$\sigma$.  Based on this result, we report only the result for the 2-parameter modified Lindhard model.

\subsubsection{\label{Uncertainties}Calculation of uncertainties }

We took into account the statistical uncertainties due to the finite size of the simulated neutron spectrum and of the experimental neutron-ON and neutron-OFF spectra by performing fits to simulated experiments. The distributions of the fit results were then fit with Gaussian distributions, and the standard deviation of each Gaussian was propagated as a statistical uncertainty for the respective contribution. We propagated the uncertainty on the cut efficiency functions by repeating the analysis twice, using the efficiencies shifted by one standard deviation up and down with respect to their central values. We took the difference in the results as estimate for the uncertainty in the yield. 

The value of the Fano factor used in the energy resolution model was determined from a fit to data (predominantly electron recoils) and it was affected by statistical and systematic uncertainties. In particular, a previous measurement~\cite{Dougherty} showed that the Fano factor for nuclear recoils could be significantly higher than in electron recoils. To account for these uncertainties, the fit was repeated twice, once forcing a downward shift for $F$ compatible with its statistical uncertainty, and once forcing an upward shift to a value of 10 for nuclear recoils, which is large enough to include any measurements in literature~\cite{Fano1,Fano2}. The resulting shift in the fit results were taken as the estimate for systematic uncertainty associated with the Fano factor. In order to evaluate the uncertainty on the background model shape, described in Sec.~\ref{bkg_model}, we recalculated the yield by using as background PDF the analytical model only, without adding the residuals from the fit to the neutron-OFF data. The difference with respect to the result obtained using the standard background model was used as an estimate for this systematic uncertainty. 

We studied the systematic uncertainty arising from the limited knowledge of the neutron elastic scattering cross section used to generate the signal model. The neutron simulation was repeated using several different realizations of cross section files described in Sec.~\ref{sec:simulation}. The negative log likelihood fit was repeated for each of the simulations, and the standard deviations of the resulting distribution of the fit results were used as a systematic uncertainty. Another source of systematic uncertainty arose from the choice of the specific neutron cross section libraries, G4NDL4.6~\cite{G4NDL46}. In order to quantify this uncertainty, another simulation was generated using the LEND~\cite{LEND} neutron cross section library. The difference between the fit results obtained using these two cross section libraries was used as an estimate of this systematic uncertainty. This approach does not take into account possible correlated biases between the two packages.

\begin{table}
 \caption{Summary of the $k_\text{{low}}$ and $k_\text{{high}}$ fit results along with their statistical and systematic uncertainties} 
\begin{ruledtabular}
\begin{tabular}{c c c c}
   & Best fit value & Stat. uncertainty & Sys. uncertainty\\
 k$_{\rm low}$  & 0.040 & 0.005 & 0.008\\
 k$_{\rm high}$  & 0.142 & 0.011 & 0.026\\
\end{tabular}
 \end{ruledtabular}
\label{table_fitresults}  
\end{table}

\section{\label{results}Results}

\begin{figure}[!ht]
    \includegraphics[width=\columnwidth]{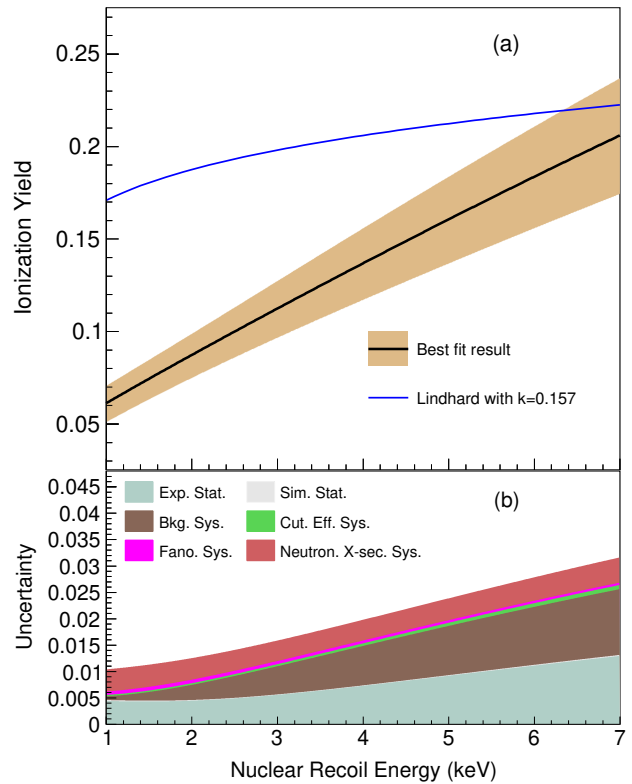}
    \caption{\label{fig:result_2param} (a) The ionization yield with 1 $\sigma$ uncertainty from the best fit values of the 2-component $k$-parameter Lindhard model as a function of the nuclear recoil energy in germanium. The blue line shows the standard Lindhard model with $k=0.157$. (b) The contribution of various sources of statistical and systematic uncertainties to the ionization yield in germanium.   The lower limit of the x-axis represents the analysis threshold converted to recoil energy using the best-fit value.}
\end{figure}

The fitted Lindhard model parameters and their respective uncertainties were propagated to the ionization yield $Y (E_{r})$ vs. nuclear recoil energy plane in Fig.~\ref{fig:result_2param}. While the resulting ionization yield was compatible with the Lindhard model with the standard $k=0.157$ value at the higher end of our energy range, our data prefer a significantly lower yield at the lower energies. As indicated by the result of the likelihood ratio test described in Sec.~\ref{likelihood}, the data were better described by a model with a linear energy variation for the $\vec{k}$ parameters, which departs from the standard Lindhard model.  Although the final result was obtained with a simultaneous fit to the 70V and 25V datasets, we performed a cross-check where we fit each set separately. The results were consistent within the statistical uncertainty.  The dominant sources of uncertainty for this measurement were the statistical uncertainty due to the size of the experimental data set,  the systematic uncertainty due to the modeling of the background as well as due to the elastic neutron-nucleus scattering cross section for germanium. This highlights the importance of developing an accurate and precise simulation for this type of analysis.
\\

\section{\label{sec:conclusions}Conclusions}

This paper presents a nuclear recoil induced ionization yield measurement for first generation, high-voltage, SuperCDMS germanium detectors operated at voltages of 25\,V and 70\,V and a temperature of approximately 50~mK. Our analysis has robustly shown that under these conditions, the ionization yield in the 1--7\,keV recoil energy region has a behavior that falls off significantly faster than the standard Lindhard model. These data also differ significantly from those of other existing measurements in this energy regime. However, previous measurements exhibit a lack of consistency among themselves, suggesting that there are systematic effects that remain to be understood within the field as a whole.  

Our model for the yield function is a first order approximation to a generic function that tends to the Lindhard functional form at high energies.  This is achieved by allowing the $k$ parameter to have a linear energy dependence in the region below 7\,keV.  To improve on this technique, future measurements could employ thinner detectors to reduce the population of multiple neutron scatters, gather a significantly larger data sample (to reduce the statistical uncertainty), and achieve a better understanding of low-energy neutron scattering cross sections implemented in the simulation (thus reducing the largest source of systematic uncertainty). 

In the case of experiments searching for the elusive scattering of DM off nuclei, the nuclear recoil ionization yield affects the signal spectrum substantially, thus injecting uncertainty into DM searches. In the low energy region, measurements in literature for germanium and silicon have not always yielded consistency with Lindhard or with each other.  Furthermore, this low-energy region is particularly important for recent dark matter searches with solid state detectors. These inconsistencies may be due to temperature, electric field, or other effects that are generally unaccounted for in measurements to date (see Fig.~\ref{fig:yield-all}). Thus, the situation suggests that multiple measurements, performed with a variety of experimental approaches and devices, are required to shed light on possible, heretofore, unquantified systematic effects in these studies.  Until those effects are clearly understood, the existing measurements will continue to be critical for determining the ionization yield under the specific operating conditions of individual dark matter experiments. The current report provides the community a new set of data via an in-situ photon-neutron measurement method, presented in comparison to the Lindhard model.

\begin{acknowledgements}

The SuperCDMS collaboration gratefully acknowledges technical assistance from the staff of the Soudan Underground Laboratory and the Minnesota Department of Natural Resources. The CDMSlite and iZIP detectors were fabricated in the Stanford Nanofabrication Facility, which is a member of the National Nanofabrication Infrastructure Network, sponsored and supported by the NSF. Funding and support were received from the National Science Foundation, the U.S. Department of Energy (DOE), Fermilab URA Visiting Scholar Grant No. 15-S-33, NSERC Canada, the Canada First Excellence Research  Fund, the Arthur B. McDonald Institute (Canada),  the Department of Atomic Energy Government of India (DAE), the Department of Science and Technology (DST, India) and the DFG (Germany) - Project No. 420484612 and under Germany’s Excellence Strategy - EXC 2121 ``Quantum Universe" – 390833306.   Femilab is operated by Fermi Research Alliance, LLC,  SLAC is operated by Stanford University, and the PNNL is operated by the Battelle Memorial Institute, each for the U.S. Department of Energy under contracts DE-AC02-37407CH11359, DE-AC02-76SF00515, and DE-AC05-76RL01830, respectively.

\end{acknowledgements}

\bibliography{photoneutron.bib}
\bibliographystyle{apsrev4-2}

\end{document}